\documentstyle[twoside,fleqn,npb,epsfig]{article}
\newcommand{\ep}{\varepsilon}

\newcommand{\MSb}{$\overline{\rm MS}$ }

\newcommand{\AmS}{{\protect\the\textfont2
  A\kern-.1667em\lower.5ex\hbox{M}\kern-.125emS}}

\title{Full two-loop electroweak corrections to the pole masses of gauge bosons}

\author{F.~Jegerlehner \address{DESY, Platanenallee 6,
                        D-15738, Zeuthen, Germany},
\addtocounter{address}{-1}
    M.~Yu.~Kalmykov \addressmark
\thanks{Work was supported in part by INTAS-CERN grant No.~99-0377
    and by the Australian Research Council grant}
and
    O.~Veretin \address{Institut f\"ur Theoretische Teilchenphysik,
        Universit\"at Karlsruhe, D-76128 Karlsruhe, Germany}
}

\begin{document}
\onecolumn{
\renewcommand{\thefootnote}{\fnsymbol{footnote}}
\setlength{\baselineskip}{0.52cm}
\thispagestyle{empty}
\begin{flushleft}
DESY 02--209 \\
November 2002\\
\end{flushleft}

\setcounter{page}{0}

\vspace*{3cm}

\begin{center}
{\LARGE\bf Full two-loop electroweak corrections } \\  
\vspace{3mm}
{\LARGE\bf  to the pole masses of
gauge bosons\footnote[3]{\noindent 
        To appear in the Proceedings of ``RADCOR 2002 - Loops and Legs 2002'', 
        Kloster Banz, 2002,
        Nucl. Phys. B (Proc.Suppl.)
}}\\
\vspace*{2cm}
Fred Jegerlehner$^{\rm a}$, Mikhail Kalmykov$^{\rm a}$\footnote[4]{\noindent 
        Work was supported in part by INTAS-CERN grant No.~99-0377
        and by the Australian Research Council grant} and
Oleg Veretin$^{\rm b}$\vspace{0.5cm}\\

\vspace*{1cm}

\noindent
$^{\rm a}${\it
Deutsches Elektronen-Synchrotron DESY, Platanenallee 6, D-15738 Zeuthen,
Germany }\\
$^{\rm b}${\it
Institut f\"ur Theoretische Teilchenphysik,
        Universit\"at Karlsruhe, D-76128 Karlsruhe, Germany }
\end{center}

\vspace{5em}
\large
\vspace*{\fill}}

\newpage

\begin{abstract}
We discuss progress in SM two-loop calculations
of the pole position of the massive gauge-boson propagators.
\end{abstract}

\maketitle
\section{INTRODUCTION}
An important type of ``universal'' (process independent) corrections
relevant for complete electroweak 2-loop calculations of physical
processes are the contributions from on-shell gauge boson
self-energies which incorporate the relation between bare,
$\overline{\rm MS}$ and on-shell (pole) masses (2-loop
renormalization constant in on-shell scheme). Such calculations are
important to scrutinize theoretical uncertainties which might obscure
e.g. the indirect Higgs mass bounds obtained by the LEP
experiments. Most interesting are $2
\rightarrow 2$ fermion processes, which in future eventually may be
investigated with much higher precision at TESLA in case the
Giga$Z$~\cite{GigaZ} option would be realized. Here, we outline
the complete 2-loop SM calculation of the pole position of the
massive gauge-boson propagators presented in~\cite{I,II}.

The position of the pole $s_P$ and the wave-function renormalization
constant $Z_2$ of the propagator of a massive gauge-boson are defined
by the relation
\begin{eqnarray}
\frac{1}{s - m_0^2 - \Pi(p^2,m_0^2,\cdots)} \simeq \frac{Z_2}{s - M^2
+ i M \Gamma}
\end{eqnarray}
for $s \simeq s_P$
where $\Pi(p^2,\cdots)$ is the transversal part of the one-particle
irreducible self-energy and $m_0^2$ is the bare mass and we have
adopted the standard parameterization of the pole
$s_{P,V} = M^2_V - i M_V \Gamma_V$, where $\Gamma_V$ is the width of
the unstable gauge-boson.

This equation can be solved perturbatively, order by order. Up to two
loops the solution reads
\begin{eqnarray}
s_P & = & m^2 + \Pi^{(1)} + \Pi^{(2)}
+ \Pi^{(1)} \Pi^{(1)}{}' \; ,
\label{polemass}
\\
Z_2^{-1} & = & 1 - \Pi^{(1)}{}'
- \Pi^{(2)}{}' - \Pi^{(1)} \Pi^{(1)}{}'' \;,
\end{eqnarray}
where $\Pi^{(L)}$ is the bare or {\MSb}-renormalized $L$-loop
contribution to $\Pi$, the prime (double prime) denotes the derivative
(second derivative) with respect to $p^2$ at $p^2=m^2$. One of the
remarkable properties of (\ref{polemass}) is that the pole position is
well-defined in terms of self-energy diagrams and its derivatives at
momentum (square) equal to the bare or the {\MSb} mass (square)
which, by construction, are real parameters.

For the $Z$-boson the equation for the position of the pole is
modified due to $\gamma-Z$ mixing $$ s_P - m_Z^2 - \Pi_{ZZ}(s_P) -
\frac{\Pi^2_{\gamma Z}(s_P)}{s_P-\Pi_{\gamma \gamma}(s_P)} = 0\;. $$
The 2-loop wave-function renormalization constant in this case is
equal to $$ Z_{ZZ}^{-1} = 1 - \Pi_{ZZ}^{(1)}{}'- \Pi_{ZZ}^{(2)}{}' -
\Pi_{ZZ}^{(1)} \Pi_{ZZ}^{(1)}{}'' - \frac{2}{m_Z^2} \Pi_{\gamma
Z}^{(1)} \Pi_{\gamma Z}^{(1)}{}'\;. $$
We would like to stress, that in order to obtain gauge invariant results
the tadpole contributions have been included~\cite{FJ}.
\section{CALCULATIONS}
At the 2-loop level we have about 1000 1PI diagrams each for the $Z$-
and the $W$-propagator. Our calculation is largely automatized: we use
${\bf QGRAF}$~\cite{qgraf} to generate the diagrams and then the
C-program {\bf DIANA}~\cite{diana} to produce for each diagram the
input suitable for our {\bf FORM} packages.

Two-loop propagator type diagrams with several masses can be reduced
to a restricted set of so-called master-integrals by using a complete
set of recurrence relations given in~\cite{tarasov-propagator}.  We
have used this approach only for the calculation of the 2-loop
massless fermion corrections, where analytical results for the
master-integrals are available~\cite{2loop-analytic-b}. As compared to
the existing calculation of the massless fermion contribution,
performed in~\cite{2loop-fermion}, we apply Tarasov's recurrence
relations~\cite{tarasov-propagator} which allows us to reduce the
number of master-integrals to a minimal set.  The latter includes
integrals which may be evaluated by using the package {\bf
ON-SHELL2}~\cite{onshell2} plus the following new prototypes: (using
standard notation~\cite{I,II})\\[-5mm]
\begin{itemize}
\item {\bf $ZZ$:}\\[-7mm]
\end{itemize}
$F_{W0W00}, F_{0000W}, V_{H00Z}, V_{W00W}, J_{00\{H,W\}}\;.$\\[-5mm]
\begin{itemize}
\item {\bf $WW$:}\\[-7mm]
\end{itemize}
$F_{Z0W00}, F_{0000Z}, V_{W00Z}, V_{\{H,Z\}00W}, J_{00\{H,Z\}}\;.$\\[3mm]
We have worked out an independent analytical calculation of these
master integrals by using a method developed in~\cite{II}.  One of the
crucial points of our approach is the observation~\cite{series} that
the analytical results for diagrams with two-massive cuts have a very
simple form if written in terms of new variables. For diagrams
with two equal internal masses $m^2$ and arbitrary external momentum
$p^2$ the new variable is $y =
\frac{1-\sqrt{\frac{z}{z-4}}}{1+\sqrt{\frac{z}{z-4}}}$, where $z =
p^2/m^2$. When the external momentum belongs to one of the
internal masses $p^2=m^2$, the new variable is $\chi =
\frac{1-\sqrt{1-4x}}{1+\sqrt{1-4x}}$, where $x = m^2/M^2$.  To speed
up the numerical evaluation of on-shell gauge-boson self-energies, an
expansion in $\sin^2 \theta_W$ may be used, which converges well and
simplifies the results considerably.  However, it is not a naive Taylor
expansion due to the presence of threshold singularities of the form
$\ln^j \sin^2 \theta_W$. We therefore briefly outline the
corresponding expansions for the relevant master-integrals by giving
two examples. The integral $F_{W0W00}~[z=1/\cos^2
\theta_W]$ depends on the small parameter $ s \equiv 1- 1/z$ via the variable
(see~\cite{II})
\begin{eqnarray}
&& \hspace{-8mm}
y = \frac{1}{1-s} \left(
\frac{1}{2} - s + {\rm i} \frac{\sqrt{3}}{2} \sqrt{1-\frac{4}{3} s}
                     \right)
= \left(\sum_{j=0} s^j \right)
\nonumber \\ && \hspace{-8mm}
\Biggl[\exp \left( i\frac{\pi}{3} \right) - s
- s \frac{{\rm i}}{\sqrt{3}}
\sum_{n=0} \left( 2n \atop n \right)  \left(\frac{s}{3} \right)^n \frac{1}{n+1}
\Biggr ]
\nonumber
\end{eqnarray}
which we expand first
and then substitute it into the analytical result in order to get the
series expansion for $F_{W0W00}$. The integral $F_{Z0W00} [x=\cos^2
\theta_W] $ depends on $s$ via the variable
\begin{eqnarray}
&& \hspace{-8mm}
\chi = \frac{1}{1-s} \left(
- \frac{1}{2} + s + {\rm i} \frac{\sqrt{3}}{2} \sqrt{1-\frac{4}{3} s}
                     \right)
=  \left(\sum_{j=0} s^j \right)
\nonumber \\ && \hspace{-8mm}
\Biggl[ \exp \left( i\frac{2 \pi}{3} \right) + s
- s \frac{{\rm i}}{\sqrt{3}}
\sum_{n=0} \left( 2n \atop n \right)  \left(\frac{s}{3} \right)^n \frac{1}{n+1}
\Biggr ] 
\nonumber
\end{eqnarray}
and we proceed as in the previous case.
However, in all other cases, the master-integrals which show up are
not know analytically so far. They can be calculated numerically,
by using one of the approaches and/or programs developed for the
2-loop self-energies~\cite{numeric}.  The full list
of master-integrals occurring in the 2-loop calculation of the pole-mass
of the gauge-bosons (for the photon propagator see~\cite{FTT00}) reads:
\begin{itemize}
\item {\bf $ZZ$:}\\[-7mm]
\end{itemize}
$F_{WWWWH},
F_{ttttH},
F_{ZWHWZ},
F_{ZHHZZ},
F_{ZZHHH},$
$F_{WWWWZ},
F_{ZtHtt},
F_{W0W0t},
F_{WtWt0},
F_{t0t0W},$
$V_{WZWW} ,
V_{WHWW} ,
V_{HWWZ} ,
V_{HttZ}  ,
V_{ZWWH} ,$
$V_{ZttH}  ,
V_{tHtt}  ,
V_{W0tW}  ,
V_{t0Wt}  ,
V_{tZtt}  ,
V_{HHZZ} ,
V_{ZZZH} ,$
$J_{ZWW}  ,
J_{HWW}  ,
J_{0WW}  ,
J_{0tt}  ,
J_{Ztt}  ,
J_{Htt}  ,
J_{0WZ}  ,$
$J_{ZHH}  ,
J_{0Wt}   \; .$
\begin{itemize}
\item {\bf $WW$:}\\[-7mm]
\end{itemize}
$F_{00ttZ} ,
F_{W0Ztt} ,
F_{W0Htt} ,
F_{WWZZH} ,
F_{WWHZZ} ,$
$F_{WWHHH} ,
F_{WZZWW} ,
F_{WZHWW} ,
F_{WHHWW} ,$
$F_{WtZ00} ,
V_{00Wt}  ,
V_{0Ztt}  ,
V_{0Htt}  ,
V_{WHHH}  ,
V_{WWWZ}  ,$
$V_{WWWH}  ,
V_{WttZ}  ,
V_{WttH}  ,
V_{WZZH}  ,
V_{WZHZ}  ,$
$V_{Z0tW}  ,
V_{H0tW}  ,
V_{ZWHW}  ,
V_{HWZW}  ,
V_{ZWZW}  ,$
$V_{HWHW}  ,
J_{0WZ}   ,
J_{0Wt}   ,
J_{0WH}   ,
J_{0Zt}   ,
J_{0Ht}   ,
J_{WHH}   ,$
$J_{WZZ}   ,
J_{Wtt}   ,
J_{WZH}   \; .$\\[3mm]
To keep control of gauge invariance we adopt the $R_\xi$ gauge with
three different gauge parameters $\xi_W,\,\xi_Z$ and $\xi_\gamma$.  In
most cases exact analytic results in terms of known functions are not
available. Thus, instead of working with the exact formulae (which only
can be evaluated numerically, at present) we resort to
some approximations, namely, we perform appropriate series expansions
in (small) mass ratios~\cite{asymptotic}.  For diagrams with several
different masses it is possible that several small parameters are
available.  In this case we apply different asymptotic expansions
(see~\cite{2region}) one after the other. Specifically, we expand in the
gauge parameters about $\xi_i=1$, in $\sin^2
\theta_W$ and, for diagrams with Higgs or/and top-quark lines, in
$m_V^2/m_H^2$ or/and $m_V^2/m_t^2$. Numerical results are
obtained using the packages {\bf ON-SHELL2}~\cite{onshell2} and {\bf
TLAMM}~\cite{tlamm} (see~\cite{I,II}).

Renormalizing the pole-mass at the 2-loop level requires to
calculate the 1-loop renormalization constants for all physical
parameters (charge and masses), as well as the 2-loop
mass-renormalization constant itself. Not needed are the
wave-function and ghost sector renormalizations.

The full 2-loop relation between pole and \MSb parameters can be
written in the form\\[-4mm]
\begin{eqnarray}
&& \hspace{-5mm}
s_{P,V} =  m_{V,0}^2 +
  \Pi^{(1)}_{V,0} + \Pi_{V,0}^{(2)}
+ \Pi^{(1)}_{V,0} \Pi_{V,0}^{(1)}{}'
\nonumber\\ && \hspace{-5mm}
+ \Biggl[ \sum\limits_j (\delta m^2_{j,0})^{(1)} \frac{\partial}{\partial m_{j,0}^2}
+ (\delta e_0)^{(1)} \frac{\partial}{\partial e_0}
\Biggr] \Pi_{V,0}^{(1)} ,
\label{mct}
\end{eqnarray}
where the sum runs over all species of particles, $(\delta e_0)^{(1)}$
and $(\delta m^2_{j,0})^{(1)}$ are the 1-loop counter-terms for the
charge and physical masses in the {\MSb}-scheme. The derivatives in
(\ref{mct}) correspond to the subtraction of sub-divergencies. The
genuine 2-loop mass counter-term is obtained by expanding
$m_{V,0}^2$ in terms of the renormalized mass.

Summing all 2-loop contributions we restore gauge invariance of the
position of the complex pole before UV
renormalization.  After UV-renormalization the propagator pole is
represented in terms of finite amplitudes.  For explicite results we
refer to~\cite{I,II}. For the numerical evaluation we were using the
first six coefficients in the weak angle $\sin^2\theta_W$ and in the
mass ratios $m_V^2/m_{heavy}^2$.

\section{\protect RENORMALIZATION~~GROUP~~EQUATIONS}

We now consider the SM renormalization group (RG) equations. We will
denote on-shell masses by capital $M$ and \MSb masses by lowercase $m$.
We adopt the following definitions for the RG functions: for all
dimensionless coupling constants, like $g,g',g_s,e,\lambda$, the
$\beta$-function is given by $\mu^2
\frac{\partial}{\partial \mu^2} g = \beta_g$ and for all mass
parameters (a mass or the Higgs v.e.v. $v$) the anomalous dimension
$\gamma_{m^2}$ is given by $\mu^2
\frac{\partial}{\partial \mu^2} \ln m^2 = \gamma_{m^2}.$ Using the
fact that $s_P$ is RG-invariant: $\mu^2 \frac{d}{d \mu^2} s_P
\equiv 0$, we are able to calculate the anomalous dimension of the
gauge bosons masses from our finite results. At the same time, a typical
relation between bare- and \MSb-masses has the form\\[-3mm]
$$
m_{V,0}^2 =  m_V^2(\mu)\:
( 1 + \sum_{k=1} Z_V^{(k)} \ep^{-k}
)$$

\vspace*{-3mm}

\noindent
such that the RG functions may be calculated directly from the UV
counter-terms $\gamma_V = \sum_j \frac{1}{2} g_j
\frac{\partial}{\partial g_j } Z_V^{(1)}$, ($j=g,g_s$). In addition, the
UV counter-terms satisfy relations connecting the higher order poles
with the lower order ones:

\vspace{-4mm}
\begin{eqnarray}
&& \hspace{-5mm}
\biggl(
\gamma_V +\sum_j \beta_{g_j} \frac{\partial}{\partial g_j }
+ \sum_i \gamma_i m_i^2 \frac{\partial}{\partial m_i^2} \biggr) Z_V^{(n)}
\nonumber \\ && \hspace{-5mm}
=  \frac{1}{2} \sum_j g_j \frac{\partial}{\partial g_j } Z_V^{(n+1)}.
\end{eqnarray}

\vspace*{-3mm}

\noindent
In the SM it is interesting to compare the RG equations calculated in
broken phase with the ones obtained in the unbroken phase. Let us
remind that at the tree-level the vacuum expectation value $v$ is
given by $ v^2 \equiv \frac{m^2}{\lambda}$, where $m^2$ and $\lambda$
are the parameters of the symmetric scalar potential.
The masses of the gauge-bosons are equal to $ m_Z^2 = \frac{1}{4} (
g^2 + g'^2) v^2$ and $ m_W^2 = \frac{1}{4} g^2 v^2$, respectively. The
fact that these relations are RG invariant on the level of the bare
quantities implies the relations
\begin{eqnarray}
&& \hspace{-5mm}
\gamma_W - 2 \frac{\beta_g}{g} = \gamma_{m^2} - \frac{\beta_\lambda}{\lambda} \;,
\\ && \hspace{-5mm}
 \gamma_Z - \gamma_{m^2} + \frac{\beta_\lambda}{\lambda} 
= 2  \left( \cos^2 \theta_W \frac{\beta_g}{g} 
+ \sin^2 \theta_W \frac{\beta_{g'}}{g'} \right),\nonumber \\
\label{rg2}
\end{eqnarray}
where the 2-loop RG functions $\beta_g, \beta_{g'}, \beta_\lambda,
\gamma_{m^2}$ have been calculated in the unbroken phase
in~\cite{RG_2loop}.
We have verified in the \MSb scheme, that these relations are valid up
to 2-loop order in the broken phase with the same RG functions. Thus
the RG equations for the \MSb masses in the broken theory can be written
as
\begin{eqnarray}
&& \hspace{-5mm}
m_W^2 (\mu^2) = \frac{1}{4} \frac{g^2(\mu^2)}{\lambda(\mu^2)} m^2(\mu^2) \;,
\nonumber \\ && \hspace{-5mm}
m_Z^2 (\mu^2) = \frac{1}{4}  \frac{g^2(\mu^2) + g'(\mu^2)}{\lambda(\mu^2)} m^2(\mu^2) \;,
\nonumber \\ && \hspace{-5mm}
m_H^2 (\mu^2) = 2 m^2(\mu^2) \;,
\nonumber \\ && \hspace{-5mm}
m_t^2 (\mu^2) = \frac{1}{2} \frac{y_t^2(\mu^2)}{\lambda(\mu^2)} m^2(\mu^2) \;,
\end{eqnarray}
where $y_t$ is the top-quark Yukawa coupling (the other
Yukawa couplings are kept zero).
The \MSb  Fermi constant
\begin{equation}
\hat{G}_F(\mu^2) \equiv \frac{ \sqrt{2} \; e^2(\mu^2) } {8 m_W^2(\mu^2) \sin^2 \theta_W(\mu^2)}
\end{equation}
satisfies the following RG equation
\begin{eqnarray}
\mu^2 \frac{\partial}{\partial \mu^2}  \ln \hat{G}_F(\mu^2)  =
\frac{\beta_\lambda}{\lambda} - \gamma_{m^2} \; .
\end{eqnarray}

The knowledge of the anomalous dimensions $\gamma_V$ allow us to write
expression for the pole positions $s_{P,V}$ with explicit factorization of
the RG logarithms. What we get is
\vspace{-0mm}
\begin{eqnarray}
&& \hspace{-5mm}
s_{P,V} =
m_V^2 \left( 1  - g^2 \gamma_V^{(1)} L_a \right)
+  g^2  X_{V,1}
\nonumber \\ && \hspace{-5mm}
+ g^4 m_V^2 \left( C^{(2,2)} L_a^2 - C^{(2,1)} L_a \right)
+ g^4  X_{V,2} \; ,
\label{factor}
\end{eqnarray}
where
$\mu^2 \frac{\partial}{\partial \mu^2} \ln m_a^2 =
\gamma_a^{(1)} g^2 + \gamma_a^{(2)} g^4$,
$L_a = \ln \frac{\mu^2}{m^2_a}$,
\begin{eqnarray}
&& \hspace{-5mm}
2 C_V^{(2,2)} = \Biggl[ \theta
+ \sum_j \gamma^{(1)}_{m_j} m_j^2 \frac{\partial}{ \partial m_j^2} \Biggr]
\gamma_V^{(1)}  \;,
\\ && \hspace{-5mm}
C_V^{(2,1)} =
\gamma_V^{(2)}
+ \gamma_V^{(1)}\gamma_a^{(1)}
\nonumber \\ && \hspace{-5mm}
+ \frac{1}{m_V^2} \Biggl[ 2 \beta_g^{(1)}
+ \sum_j \gamma^{(1)}_{m_j} m_j^2  \frac{\partial}{ \partial m_j^2} \Biggr] X_{V,1} \;,
\end{eqnarray}
and $\theta = \gamma_V^{(1)} + 2 \beta_g^{(1)}$, $\mu^2 \frac{\partial
}{\partial \mu^2} g = \beta_g^{(1)} g^3$.  In contrast to QCD,
$\gamma_V^{(1)}$ and $C_V^{(i,j)}$ have non-polynomial structure in
the massless coupling constants which originated from the tadpole
contributions. $X_{V,1}$ has been calculated long time ago~\cite{FJ}
and $X_{V,2}$ are our results which, for the numerical evaluation, we
approximated by finite series. The functions $X_{V,j}$ we have
represented in terms of $\overline{\rm MS}$ parameters.  We note that
the amplitudes $X_{V,j}$ entering (\ref{factor}) have no
explicit $\mu$ dependence.

\section{SCHEME DEPENDENCE}

Our results reveal terms of unexpectedly high powers of the Higgs and
the top-quark masses in (\ref{factor}), arising from 2-loop
corrections.  In fact the purely bosonic diagrams yield $m_H^4/m_V^4$
terms and the $(t,b)$ quark-doublet ($m_t \gg m_b$) contributes
$m_t^6/(m_H^2 m_V^4)$ power corrections.  At a first glance, such
terms contradict Veltman's screening
theorem~\cite{Veltman:screening} which states, that the
L-loop Higgs dependence of a physical observable is bounded by
$(m_H^2)^{L-1} \ln^L m_H^2 $ for large Higgs masses.  However, this
theorem only applies to physical observables like cross sections and
asymmetries. If we consider quantities like $\Delta r$ (which is an
observable in the on-shell scheme) in the $\overline{\rm MS}$ scheme
the screening theorem does not hold in general. To illustrate this,
let us compare the 1-loop EW corrections to the Fermi
constant~\cite{muon} in the on-shell and the $\overline{\rm MS}$
scheme\footnote{We should mention that $\Delta r^{(1)}_{MS}$
introduced in (\ref{MS:muon}) is different from $\Delta \hat{r}$
defined via a hybrid scheme (couplings \MSb, masses on-shell)
in~\cite{muon:MS}.}:
\vspace{-1mm}
\begin{eqnarray}
G_F & = & \frac{\pi \alpha}{\sqrt{2}} \frac{1}{M_W^2 \sin^2 \Theta_W}
\left( 1+\Delta r^{(1)}_{OS} \right)
\nonumber \\
    & = &
\hat{G}_F(\mu^2) \left( 1+\Delta r^{(1)}_{MS} \right) 
\label{MS:muon}
\end{eqnarray}
so that
$$
\Delta r^{(1)}_{MS} = \Delta r^{(1)}_{OS}
+ \Biggl[\frac{m_W^2}{M_W^2} - 1 \Biggr] + \Biggl[\frac{\sin^2
\theta_W}{\sin^2 \Theta_W} - 1 \Biggr] + \Delta \alpha \; .  $$  
The 1-loop correction $\Delta r^{(1)}_{OS}$ in the limit of a heavy
Higgs boson has a logarithmic Higgs mass dependence only . In
contrast, in the \MSb scheme, higher powers of the Higgs mass are
showing up, because $\frac{m_W^2}{M_W^2} - 1$ contributes the term $$
\Delta r^{(1)}_{MS} \sim - \frac{\sqrt{2} \hat{G}_F(\mu^2) m_W^2}{16 \pi^2}
\frac{7}{2}  \frac{m_H^2}{m_W^2} + O(m_H^2) \; ,
$$ while all other corrections exhibit logarithmic behavior in the
Higgs mass, only\footnote{The hybrid quantity $\Delta \hat{r}^{(1)}$
does not show extra powers of the Higgs mass since masses are kept
on-shell.}.  Our relation (\ref{factor}) between parameters of two
different schemes does {\em not} relate physical observable but it is
very important for analysis of the uncertainties coming from higher
order effects~\cite{scheme}.  In particular, our estimations of the
2-loop boson contribution to $\Delta r$~\cite{I} is very close to
the results of real calculations~\cite{r-boson-1,r-boson-2}.
\section{CONCLUSION}
By an explicite 2-loop calculation we have shown that (to this order):
${\bf 1.}$ The position of the complex pole $s_p$ of a the gauge-boson
($Z,W$) propagator is a gauge invariant quantity after inclusion of
the Higgs tadpole contributions (see also~\cite{pole:SM}).  ${\bf 2.}$
The renormalized on-shell self-energies are infrared finite. This
derives from the fact that within dimensional regularization, which
allows to regularize UV and IR singularities by the same $\ep$ ($\ep
=(4-d)/2\to0$) parameter, the singular $1/\ep$ terms are absent after
UV renormalization.  ${\bf 3.}$ The inclusion of the tadpoles is
important for the renormalization group invariance and for the gauge
invariance of the parameter renormalizations.  ${\bf 4.}$ By our
calculation we have proven that the \MSb renormalization scheme is
self consistent and works properly in case of unstable particles.
${\bf 5.}$ Our results for the 2-loop mass renormalization constants
in the on-shell and the \MSb scheme can be applied in calculations
of physical quantities in both of these schemes at the 2-loop
level. Examples of such calculations, where results of our
paper~\cite{I} have been used, are the computations of the bosonic
2-loop contributions to the Muon life-time presented
in~\cite{r-boson-2}.


\end{document}